\documentclass[pre,a4paper,twocolumn,showkeys]{revtex4}
\usepackage{graphics}

\begin{document}

\title{Trends in Atomic Adsorption on Titanium Carbide and Nitride}

\author{Aleksandra Vojvodic}
\email{alevoj@fy.chalmers.se}
\author{Carlo Ruberto}
\author{Bengt I. Lundqvist}

\affiliation{Department of Applied Physics, Chalmers University of Technology, 
SE-412 96 G\"{o}teborg, Sweden}


\begin{abstract}

Extensive density-functional calculations on atomic chemisorption 
of H, B, C, N, O, F, Al, Si, P, S, and Cl on the polar TiC($111$) and 
TiN($111$) yield similar adsorption trends for the two surfaces:  
(i) pyramid-like adsorption-energy trends along the adatom periods;  
(ii) strongest adsorption for O, C, N, S, and F;  
(iii) large adsorption variety;  
(iv) record-high adsorption energy for O ($8.4$ -- $8.8$ eV).  
However, a stronger adsorption on TiN is found for elements on the left 
of the periodic table and on TiC for elements on the right.  
The results support that a concerted-coupling model, proposed for 
chemisorption on TiC, applies also to TiN.  

\end{abstract}

\maketitle

\section{Introduction}

Titanium carbonitrides, Ti(C,N), are in broad technical use, including
mechanical applications such as wear-resistant cutting-tool coatings,
and electronic ones, such as TiC as substrate for growth of SiC, graphene,
and other carbidic nanostructures, and TiN as diffusion barrier in 
integrated circuits. Also, the bonding nature of the technologically
fascinating MAX phases, like Ti$_3$SiC$_2$ and Ti$_3$AlC$_2$, relates to
the bonding between ``sheets'' of Ti$_6$C octahedra and Al/Si atoms
\cite{MAX}.
Common to processes like these is an initial chemisorption step that calls
for a fundamental understanding of atomic chemisorption on these materials.
Their intriguing combination of covalency, ionicity, and metallicity puts
TiC and TiN into a class of their own, suggesting unique chemisorption properties.
Can these be described with models used for transition-metal substrates?  
How do the activity and other chemisorption properties vary when going from
TiC to TiN or when changing the relative amounts of C and N in Ti(C,N)?
The higher chemical activity of the TiC($111$) surface, compared to
TiC($001$), has been suggested to be due to the presence of a surface state (SS) 
on TiC($111$) \cite{Oshima}.  Studies of SS's on TiN($111$) and of their 
importance for reactivity still seem to be lacking in the literature.

In a separate study, we perform an extensive trend study of atomic
chemisorption on the TiC($111$) surface, there providing a model for the
description of chemisorption \cite{Ruberto}.  
Here, we extend our study to TiN($111$): how do TiC($111$) and TiN($111$)
compare concerning atomic adsorption? In particular, how does the change of
substrate influence adsorption strength, adsorption-energy trends,
adsorption-site preference, adsorption geometry, and diffusion barriers?  
What are the implications of such results for the nature of the chemisorption 
on TiC and TiN and for their technological applications?

\section{Computational Method}

First-principles calculations based on the density-functional theory (DFT) are 
performed with the pseudopotential-plane-wave-based code {\tt dacapo}, 
with the PW91 generalized-gradient approximation \cite{dacapo}.  
Slabs with 4 TiX (X = C or N) bilayers (one bilayer corresponds to one Ti 
and one X atomic layer) are used.  Each atomic layer in TiC (TiN) 
is composed of 6 (9) atoms.  Convergence is tested by increasing 
the number of atoms per layer, the ${\bf k}$-point sampling, 
the plane-wave cutoff, the number of layers, and the vacuum 
thickness.  The adatom and the top three TiX bilayers are allowed 
to relax in all directions until forces are less than 
0.05 eV/\AA .  
Charge localization around individual atoms is 
measured with the ``atoms-in-molecule'' method of Bader \cite{Bader}.  
The adsorption energies $E_{\rm ads}$ are calculated as 
$E_{\rm ads} = |E_{\rm slab+adatom} - E_{\rm clean\ slab} - 
E_{\rm free\ adatom}|$.

\section{Brief Account of T\lowercase{i}C results}

The stable bulk crystal structure of TiC is the rocksalt structure,
with lattice constant of $4.332$ \AA\ from our structure optimization \cite{Ruberto} 
(experiment: $4.330$ \AA\ \cite{Dunand}) 
and a calculated cohesive energy of $14.75$ eV.  
The bonding \cite{Schwarz,Ruberto} 
is characterized by spatially directed Ti--C bonds that secure 
strength, structure, and hardness \cite{Jhi}.  
Covalency is manifested by the strong splitting of the electron states into
an empty conduction band (CB) of antibonding C$2p$--Ti$3d$ states above the
Fermi level $E_F$, and an occupied upper valence band (UVB) of bonding 
C$2p$--Ti$3d$ states, extending between $-6.1$ eV (all energies are given relative to 
$E_F$) and $E_F$.  The Fermi level lies in the 
pseudogap between the CB and the UVB, where the low but nonvanishing DOS gives the
metallic character.  A considerable ionic contribution to the bonds is 
inferred from the dominance of Ti-localized states in the CB and of 
C-localized states in the UVB, as well as from the charge transfer 
from Ti to C ($1.51$ electrons from our Bader analysis).  
Our state-resolved analyses of the UVB show also a dominance of direct C--C 
bonding states in the low-energy DOS peak of the UVB \cite{Ruberto}.  
A lower valence band (LVB), consisting of non-bonding 
C$2s$ states, extends between $-12.8$ and $-8.4$ eV.  

The polar TiC($111$) surface is Ti terminated, chemically very active, and
has a strong SS near the Fermi energy $E_F$ \cite{Oshima}.  
This SS consists 
of strongly localized surface-Ti $3d$ electrons that extend out into the
vacuum and to neighboring surface fcc sites, where they connect with the
electron clouds from the neighboring surface Ti atoms, avoiding the regions
corresponding to the hcp adsorption sites.  
In contrast, the more stable 
TiC($001$) surface is much less active and lacks a major SS \cite{Oshima}.
Like bulk TiC, the TiC($111$) surface has a UVB, dominated by C$2p$ states
in covalent Ti--C bonds and, in the lower UVB region, C--C bonds, and a
separated LVB of C$2s$ states.  

Our calculations \cite{Ruberto} show TiC able to produce atomic chemisorption
energies in a wide range, with great sensitivity to adsorbate species
[between
the record-high $8.8$ eV for O and the low $3.4$ eV for Al in the most stable
site (fcc) on TiC($111$)], as well as to substrate face [$5.0$ eV for O in
the on-top-C TiC($001$) site] and adsorption site [{\it e.g.}, $7.9$ and $6.5$ eV
for hcp and top O, respectively, on TiC($111$)].

On TiC($111$), our calculated adsorption energies $E_{\rm ads}$ for the
second- and third-period elements B, C, N, O, F, Al, Si, P, S, and Cl in
fcc, hcp, and top sites \cite{Ruberto} show pyramid-shaped trends along each period, 
with strongest binding for the group-VI elements (O and S) (see Fig.\ \ref{fig:E_ads} 
for the fcc energies, similar trends are obtained for hcp and top). The
adsorption strength weakens monotonically when moving away from group VI
in each period and when moving from period 2 to period 3. The exception
is for C and N, whose $E_{\rm ads}$ values lie very close. 
For all adatoms, the relative site stability is fcc $>$ hcp $>$ top.  
The bridge site is unstable for all adatoms and relaxes to the fcc site.  

Detailed analyses and comparisons of the calculated DOS's for the different
adatoms yield a model for the chemisorption mechanism that is based on the
Newns-Anderson (NA) model \cite{Newns}. In this concerted-coupling model 
\cite{Ruberto}, two different types of adatom--substrate couplings work together 
in a concerted way: (i) the adatom frontier orbitals overlap strongly with the
Fermi-level SS that sticks out from the surface, thus giving a strong
chemisorption in the NA sense, with well separated bonding (just below the
adatom level, deeper for more electronegative adatoms) and antibonding levels
(just above the SS and the Fermi level); (ii) in addition, for
several adatoms, there is a chemisorptive interaction with substrate states
in the UVB, with varying strength through the adatom period, as the 
SS-modified adlevel traces different parts of the varying TiC UVB features.

\section{Results for T\lowercase{i}N}

\subsection{Bulk structure, energetics, and bonding nature}

\begin{figure}
\scalebox{.47}{\includegraphics{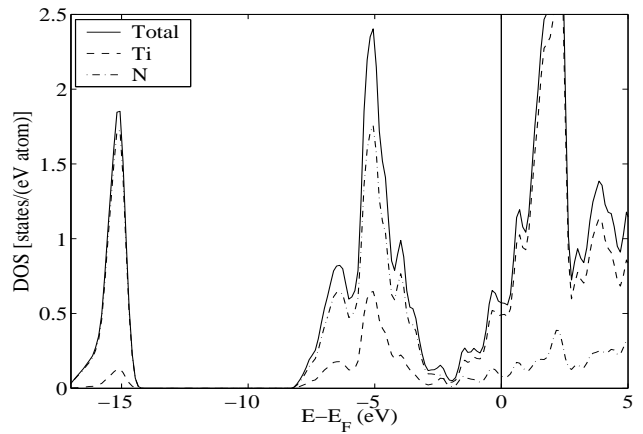}}
\caption{\label{fig:LDOS_bulk_TiN}Calculated total and atom-projected densities 
of states (DOS) for bulk TiN.}
\end{figure}

Like TiC, bulk TiN adopts the rocksalt structure.  Our structure 
optimization yields a lattice constant of $4.244$ \AA\ 
(experiment: $4.238$ \AA\ \cite{Marlo}) and a cohesive energy of $13.79$ eV.  
Compared to TiC, the electronic structure of TiN shows strong similarities, 
with a DOS (Fig.\ \ref{fig:LDOS_bulk_TiN}) consisting of bonding (in the UVB) and 
antibonding (in the CB) Ti--N states, but also differences:  
(i) a higher position of $E_F$, shifting the CB, UVB, and 
LVB to lower energies;  
(ii) a population of antibonding states in the CB, due to the 
extra electron per formula unit, which lowers the cohesion 
energy;  and 
(iii) a larger separation between UVB and CB, reflecting the 
higher ionicity of TiN (Bader charge transfer of $1.62$ electrons).

\subsection{Surface energetics and structure}

The preferred termination of TiN($111$) has only been addressed in theoretical studies, 
showing that the surface is N terminated when in an N-rich environment \cite{Gall}.  
In order to compare the stabilities of the Ti and N terminations in vacuum, 
we follow the approach used by Ref.\ \cite{Tan} to show why TiC($111$) is 
Ti terminated and compare the evaporation rates for N and Ti from TiN($111$).  
While our calculated evaporation energies for TiC($111$) 
[$E_{\rm evap}=5.40$ ($9.43$) eV/atom for C (Ti)] confirm the results of 
Ref.\ \cite{Tan}, our values for TiN($111$) [$E_{\rm evap}=6.83$ ($6.98$) eV/atom for 
N (Ti)] imply that both Ti and N are possible terminations.  
Therefore, since our focus here is to find general adsorption mechanisms and to compare 
TiN with TiC, we study the Ti termination for both surfaces.  

Our TiX($111$) slabs are stoichiometric.  Hence, our calculations yield only 
values for the sum of the surface energies of the Ti- and X-terminated 
surfaces.  This corresponds to the energy cost of cleaving the 
infinite crystal into two semi-infinite ones.  Our calculated values are:  
$E_{\rm cleav} = 9.81$ ($11.74$) J/m$^2$ for the unrelaxed TiN (TiC) surface and
$E_{\rm cleav} = 9.17$ ($11.43$) J/m$^2$ after relaxation of only the Ti-terminated 
side of the TiN (TiC) slab.  
These values should be compared with those for the nonpolar ($001$) surface, 
$2.76$ ($3.46$) J/m$^2$ for TiN (TiC) \cite{Dudiy}.  
Thus, for both TiX's, the polar ($111$) surface has a much lower stability 
than the ($001$) surface.  Also, we note that surface formation is easier on 
TiN than on TiC.  

\begin{figure}
\scalebox{.4}{\includegraphics{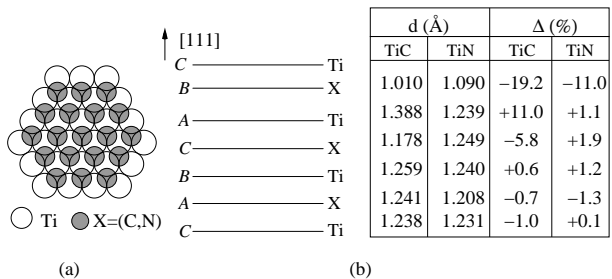}}
\caption{\label{fig:surf_struc}Surface structure of TiX($111$):  
(a) top view of the two surface atomic layers;  
(b) side view of the seven surface atomic layers, 
showing the $ABC$ stacking of alternating Ti and X layers and our 
calculated relaxations, in absolute values ($d$) and relative ($\Delta$) 
to the bulk interlayer distances ($1.251$ \AA\ for TiC and $1.225$ \AA\ 
for TiN).}
\end{figure}

Our relaxed surface structure of the Ti-terminated TiN($111$) surface 
(Fig.\ \ref{fig:surf_struc}) is in good agreement with the results of Ref.\ 
\cite{Marlo}.  The relaxation is smaller than on TiC($111$) 
and does not follow the alternating positive-negative relaxation typical of 
metals that is obtained for TiC($111$).  At the same time, it can be noted 
that the relaxations are larger than those typical for metals.

\subsection{Adsorption energetics and structure}

\begin{figure}
\scalebox{.47}{\includegraphics{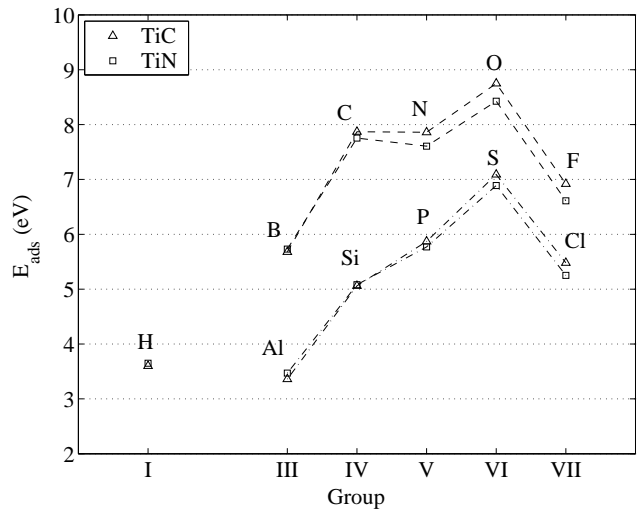}}
\caption{\label{fig:E_ads}Calculated atomic fcc adsorption energies 
$E_{\rm ads}$ on TiC($111$) and TiN($111$).}
\end{figure}

Our trend study for atomic adsorption on the Ti-terminated TiN($111$) 
comprises the same adatoms as those considered on TiC($111$), however, 
with larger focus on the second-period adatoms.  
The results for the adsorption energies $E_{\rm ads}$ 
(Fig.\ \ref{fig:E_ads} and Table \ref{tab:Eads_111}) 
show many similarities to those for TiC($111$):  
(i) at least for C, N, O, and F, a preference for the fcc site, 
followed by the hcp and top sites, while the bridge site relaxes to fcc;  
(ii) a large variety in adsorption strength, from O 
($E_{\rm ads} = 8.4$ eV in the fcc site) to Al ($3.5$ eV in the fcc site);  
(iii) overall strongest adsorption for O, C, N, S, and F;  
(iv) very pronounced pyramid-shaped $E_{\rm ads}$ trends for both 
adatom periods (with the exception of C), with strongest adsorption for 
group-VI adatoms;  
(v) a stronger adsorption for second-period adatoms.  
Also, the heights of the diffusion barriers between fcc and hcp sites can be 
estimated from the $E_{\rm ads}$ values obtained after perpendicular 
relaxation of the bridge adatoms (Table \ref{tab:Eads_111}).  
These indicate that the barriers are generally higher on TiN($111$) 
than on TiC($111$) (by $\sim 0.3$ -- $0.7$ eV for C, N, and O, and by 
$\sim 0.2$ -- $0.3$ eV for F).

\begin{table}
\caption{\label{tab:Eads_111}Calculated atomic adsorption energies $E_{\rm ads}$ 
on TiN($111$) [corresponding TiC($111$) values within parentheses].  
For all adatoms, the bridge site is unstable and relaxes to the 
neighboring fcc site.  The bridge values given correspond to 
only perpendicular relaxation.}
\vspace*{.2em}
\begin{tabular}{ccccc}
\hline\hline
 & \multicolumn{4}{c}{$E_{\rm ads}$ (eV/atom)} \\
\cline{2-5}
atom & fcc site & hcp site & top site & bridge site \\
\hline
C  & 7.75 (7.87)& 7.20 (7.15) & 4.55 (4.69)& 6.26 (6.73)\\
N  & 7.61 (7.86)& 6.97 (6.87) & 4.47 (4.59)& 6.27 (6.77)\\
O  & 8.43 (8.75)& 8.02 (7.93) & 6.28 (6.50)& 7.27 (7.95)\\
F  & 6.61 (6.92)& 6.45 (6.46) & 5.75 (6.01)& 6.05 (6.58)\\
\hline\hline
\end{tabular}
\end{table}

It is striking how closely the $E_{\rm ads}$ curves for TiC($111$) and 
TiN($111$) lie, the differences being smaller than $0.32$ eV.  
However, interesting differences arise:  
(i) the fcc adsorption is slightly stronger on TiN for the adatoms on the left side of 
each period (H, B, Al), but stronger on TiC for the adatoms on the right side 
(C, N, O, F, P, S, and Cl);  
(ii) for the fcc site, the preference for TiC adsorption increases successively when 
moving to the right within each adatom period;  
(iii) while on TiC($111$), the $E_{\rm ads}$ values for C and N are almost 
the same, on TiN($111$), N has a lower value than C.  

The two substrates show also similarities in fcc adsorption geometries 
(Table \ref{tab:ads_geom}):  
(i) the adatom--substrate bond distances are larger for period 3 
(due to the extra filled electron shell);  
(ii) the adatom--substrate distances are smallest for group V (in period 2) 
and for group VI (in period 3), and increase monotonically when moving 
away, in each period, from these (indicating a stronger adsorption 
strength for groups V--VI);  
(iii) the Ti--Ti distances around the adatoms decrease 
from the clean-surface values (indicating a strong adatom--Ti 
attraction), 
except for F and Cl, where they stay unchanged.  
In each period, the values for F and Cl stand out from the rest, due 
to their overall larger values.

\begin{table}
\caption{\label{tab:ads_geom}Calculated geometries around the fcc adatoms on 
TiN($111$) [corresponding TiC($111$) values within parentheses]:  
$d_{\rm ad-Ti}$ and $d_{\rm ad-X}$ are the distances between the adatoms and 
their nearest-neighbor Ti and X atoms, respectively;  
$Z_{\rm ad-TiX}$ is the perpendicular distance between the adatoms and the 
TiX($111$) surface;  
$d_{\rm Ti-Ti}$ is the distance between the Ti atoms closest to the adatom 
(the distances in the clean surfaces are included for comparison).  
The Ti--X and X--X distances closest to the adatom are unaffected by 
the adsorption.  All values are in \AA .}
\vspace*{.1em}
\begin{tabular}{ccccc}
\hline\hline
atom & $d_{\rm ad-Ti}$ & $d_{\rm ad-X}$ & $Z_{\rm ad-TiX}$ & 
   $d_{\rm Ti-Ti}$ \\
\hline
H  & 1.99 (2.01) & 2.71 (2.73) & 1.01 (1.03) & 2.97 (2.98) \\
\hline
B  & 2.14 (2.16) & 3.02 (3.01) & 1.33 (1.32) & 2.90 (2.95) \\
C  & 1.98 (1.99) & 2.88 (2.86) & 1.12 (1.10) & 2.84 (2.88) \\
N  & 1.94 (1.94) & 2.83 (2.80) & 1.04 (1.02) & 2.83 (2.85) \\
O  & 1.97 (1.98) & 2.85 (2.81) & 1.08 (1.07) & 2.86 (2.88) \\
F  & 2.15 (2.16) & 2.92 (2.93) & 1.28 (1.28) & 3.00 (3.00) \\
\hline
Al & 2.66 (2.67) & 3.56 (3.54) & 2.02 (2.02) & 2.99 (3.03) \\
Si & 2.53 (2.52) & 3.46 (3.39) & 1.87 (1.82) & 2.95 (3.00) \\
P  & 2.44 (2.44) & 3.37 (3.32) & 1.75 (1.72) & 2.95 (2.99) \\
S  & 2.42 (2.44) & 3.33 (3.30) & 1.73 (1.72) & 2.96 (3.00) \\
Cl & 2.51 (2.54) & 3.36 (3.35) & 1.81 (1.82) & 3.01 (3.06) \\
\hline
Clean surface 
   &     ---     &    ---      &    ---      & 3.00 (3.06) \\
\hline\hline
\end{tabular}
\end{table}

\section{Discussion and Conclusions}

Our results for atomic chemisorption on TiC($111$) and TiN($111$) show 
strong similarities.  
Since the high chemical activity of TiC($111$) has been connected to the presence of a 
SS, it is plausible to believe that, despite the electron-structure 
differences, the similar results for TiN($111$) are caused by the presence of a 
similar SS on this surface.  
The preference for adsorption in fcc site confirms this, 
considering that the TiC($111$) SS extends toward the fcc 
sites \cite{Ruberto}.  

However, the differences raise intriguing questions:  why is 
adsorption stronger on TiN on the left of the periodic table but stronger on TiC 
on the right?  
Also, why does the C adatom not follow the pyramid-shaped trend and 
instead adsorb more strongly than N?  
Obviously, factors beyond the $d$-band model \cite{Hammer}, which has been so successful 
for transition metals, have to be introduced.  
Our previous results for adsorption on TiC($111$) 
show the presence of two chemisorption mechanisms, involving both 
the SS and the UVB \cite{Ruberto}.  Since the UVB energy is shifted downwards 
in TiN, our concerted-coupling model could provide the answer to the 
$E_{\rm ads}$ differences between TiC and TiN.  This is the subject 
of a separate conference contribution \cite{SS2}.  

\begin{figure}
\scalebox{.4}{\includegraphics{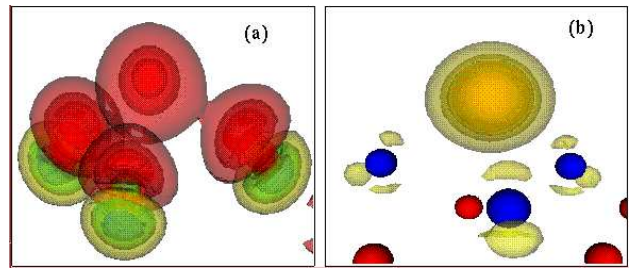}}
\caption{\label{fig:chd}Calculated differences in electron charge density for 
(a) O and (b) F adsorbed on TiN($111$) before and after adsorption.  
The color coding ranges from green (electron charge depletion due to 
adsorption) through yellow (no charge difference) to red 
(electron charge increase).  Blue (red) balls: Ti (N) atoms.}
\end{figure}

Also, the larger bond distances for F and Cl suggest a different bonding 
mechanism for these adatoms.  Indeed, the differences in charge density due to 
adsorption (Fig.\ \ref{fig:chd} shows O and F) show very weak distortions of the 
adatom-centered density for F, indicating a very high ionic character for this 
bond.  In contrast, for O, the density distorts toward neighboring Ti atoms in 
a typical covalent fashion.

Our results might be relevant for the technological applications.  
The strong O adsorption indicates that TiC($111$) and TiN($111$) 
are good substrates for oxide-layer growth.  
Indeed, both are used in multilayer CVD alumina coatings on wear-resistant 
cutting tools.  Our results suggest that nucleation and initial growth are similar 
on the two surfaces.  
For example, the higher adsorption energy for O than for Al suggests that the 
first atomic layer of the alumina coating consists of O atoms.  Also, the 
high energy for S suggests it to be a strong competing candidate, which 
can explain the presence of S found in the alumina coatings under certain conditions, 
affecting the alumina phase composition \cite{Halvarsson}.  
Thus, the preference for one or another of the two substrates is more likely due 
to considerations related to other factors, such as mechanical properties 
and lattice mismatch.  
Also, it is likely that there are differences in electronic structure between 
the two O/TiX($111$) systems, with implications for the continuing 
growth of the coating.  
In another context, the easy formation of a titanium-oxide layer on TiN may improve its 
biocompatibility \cite{Huang}.  

Although proper bulk and growth calculations obviously are needed, 
our calculated diffusion-barrier estimates might be of interest for 
the synthesis and properties of the MAX phases.  Their good 
plasticity is related to a weak bonding between the ($111$) face 
of the Ti$_6$C sheets and the A component ({\it e.g.}, Al or Si) \cite{MAX}.  
Our calculated $E_{\rm ads}$ differences between fcc, hcp, and bridge sites for 
Al and Si on TiC($111$), indicating diffusion barriers of the order of 
$0.1$ -- $0.3$ eV (activation temperatures of $\sim 50$ -- $130$ K) \cite{Bogicevic}, 
suggest good lateral mobility between the Al/Si and Ti$_6$C sheets.  
Also, our calculations indicate higher diffusion barriers on TiN($111$), 
suggesting a decrease in plasticity when substituting C with N.  

Obviously, more studies addressing these questions are highly desirable.  
As a first step, we pursue in a separate publication a detailed comparison between 
TiC and TiN of the adsorbate electronic structures, showing the applicability of the 
concerted-coupling model for adsorption on TiN($111$) \cite{SS2}.


\end{document}